\documentclass[11pt]{article}
\usepackage{a4wide}
\usepackage{authblk}
\usepackage{graphicx,xcolor}

\usepackage{caption}
\usepackage{subcaption}

\usepackage{color}
\usepackage{amsmath,amsfonts,amssymb,graphicx,hyperref,hypcap}
\usepackage{epsfig}
\usepackage{amsmath,amsfonts,epsf}
\usepackage{amssymb,tikz}

\newcommand{\done}[2]{\frac{d{#1}}{d{#2}}}

\begin{document}

\begin{center}
{\Large\bf
Diffusive dark matter and dark energy scenario and $k-$essence in the context of Supernova Ia observations}
\end{center}
\vspace{4mm}
\begin{center}
{\large Abhijit Bandyopadhyay\footnote{Email: abhijit@rkmvu.ac.in} and Anirban Chatterjee\footnote{Email: anirban.chatterjee@rkmvu.ac.in}}
\end{center}

\begin{center}
Department of Physics\\Ramakrishna Mission Vivekananda University\\
Belur Math, Howrah 711202, India
\end{center}
\vspace{4mm}

\begin{abstract}
We consider a unified model of interacting dark matter and dark energy 
to account for coincidence of
present day dark energy and dark matter densities.
We assume dark energy to be represented by a homogeneous scalar field $\phi$ 
whose dynamics is driven by a (non-canonical)
$k$-essence Lagrangian with constant potential and the particles of dark matter 
fluid undergoing velocity
diffusion in  background medium of the $k-$essence scalar
field $\phi$.
This results in a transfer of energy from the fluid of dark matter to 
that of dark energy. 
This effect  shows up as a source term
in the continuity equation for dark matter and dark energy fluids.
The source term involves a diffusion coefficient which is a measure of 
average energy transferred per unit time due to diffusion.
We use time evolutions of the scale factor of background FRW spacetime,
energy density and pressure of the dark fluid obtained from analysis of 
Supernova Ia data to obtain bounds on the diffusion parameter.
For a constant potential in the $k$-essence Lagrangian,
the temporal behaviour of a homogeneous  $k$-essence field $\phi$ is obtained
for different values of the diffusion parameter.
The obtained temporal behaviour may be expressed as 
$\phi(t/t_0) = \phi_0 + \varepsilon_1 (t/t_0 - 1) + \varepsilon_2 (t/t_0 -1)^2$,
where $t_0$ is the time corresponding to present epoch.
The coefficients $\varepsilon_1$ and $\varepsilon_2$ have been found
and obtained as  linear functions of  diffusion parameter.
\end{abstract}

\section{Introduction}
\label{sec:intro}
Measurement of luminosity distance of the type Ia Supernovae (SNe Ia) is 
the key observational ingredient in establishing the fact the universe has 
undergone a transition form decelerated to accelerated phase of expansion.
This was first reported in 1998 
independently by  
Riess \textit{et. al.} \cite{ref:Riess98}
and Perlmutter \textit{et. al.}  \cite{ref:Perlmutter}.
Source of this late time cosmic acceleration 
is generally labelled as 
`Dark Energy' (DE).
Besides, observation of rotation curves of
spiral galaxies \cite{Sofue:2000jx}, gravitational 
lensing \cite{Bartelmann:1999yn},  
Bullet cluster \cite{Clowe:2003tk}
and other colliding  clusters
provide evidence for   non-luminous 
matter in present universe  
manifesting its existence through gravitational interactions.
Such `matter', different from baryonic matter, 
is generally termed as `Dark Matter' (DM).
Measurements in satellite 
borne experiments - WMAP \cite{Hinshaw:2012aka} and 
Planck \cite{Ade:2013zuv}
established that,  dark energy
and dark matter jointly contribute  around  96\% 
($\sim 69\%$ dark energy and $\sim 27\%$ dark matter) 
of total energy density
of  present universe. Rest 4\% contribution
comes from baryonic matter with negligible contribution from radiations.
Despite countless  searches for a physical
theory of dark energy and dark matter, their origin   
and nature  still remain  a mystery.
However, the  $\Lambda-$CDM model 
\cite{weinberg-cdm} provides 
an  excellent agreement with a wide variety of
cosmological data. Here `CDM' denotes
Cold Dark Matter content of the universe and $\Lambda$,
the cosmological constant,  represents 
dark energy of vacuum from viewpoint of particle physics.
One of the major drawbacks of the model is the
large disagreement between 
estimated energy density of vacuum 
and observed value of dark energy density.
Also the same order of magnitude of observable values of dark matter
and dark energy densities at present epoch seems  accidental.
Out of diverse theoretical attempts  to resolve above
mentioned problems, one interesting approach addresses the above two
issues by considering a unified model of dark matter 
and dark energy 
\cite{Szydlowski, calogero, calogero1, Haba:2016swv}
where a dynamical relation between
dark matter and dark energy was suggested to account for
coincidence of present day dark energy and dark matter densities.\\

We assume dark energy to be represented by a homogeneous scalar field $\phi$ 
whose dynamics is driven by a
$k$-essence Lagrangian with constant potential and the particles of dark matter 
fluid undergo velocity
diffusion in  background medium of the $k-$essence scalar
field $\phi$.
This results in a transfer of energy from the fluid of dark matter to 
that of dark energy. 
This effect  shows up in the form of a non-zero source term
in the continuity equation for the fluid of dark matter and dark energy as well where
the source term involves a diffusion coefficient parameter which is a measure of 
average energy transferred per unit time due to diffusion.
The continuity equation for dark matter fluid connects
time evolution of dark matter energy density with the diffusion coefficient.
Using luminosity distance and redshift measurements from observations of Supernova Ia (SNe Ia) 
we obtain temporal behaviour of certain cosmological parameters: scale factor $a(t)$ corresponding to
Friedman-Robertson-Walker (FRW) spacetime background,  total energy density and pressure of the dark fluid
during late time phase of cosmic evolution. 
Using the results extracted from
the  SNe Ia  data and the measured value of energy density of dark matter
at present epoch from WMAP \cite{Hinshaw:2012aka} and 
Planck \cite{Ade:2013zuv} experiments, in the continuity equation for dark matter fluid 
we obtain the bounds on the diffusion coefficient parameter.
We also obtain individual time dependences of energy densities of
dark matter and dark energy components
at different values of the diffusion parameter. \\

The dynamics $k-$essence scalar field $\phi$ playing the role of background
medium in which diffusion takes place is governed by
non-canonical Lagrangian of the form $L=V(\phi)F(X)$,
where $X=(1/2)g^{\mu\nu}\nabla_\mu\phi \nabla_\nu\phi$ and 
the potential $V(\phi)$ is assumed constant. The constant potential
in $k-$essence Lagrangian ensures a scaling relation 
$X(dF/dX)^2=Ca^{-6}$ where $C$ is a constant. We identify 
the stress energy tensor corresponding to this Lagrangian
with that of the dark energy fluid. Considering the scalar field $\phi$ to 
be homogeneous and exploiting the scaling relation we may directly 
relate the time derivative of scalar field with 
energy density and pressure of dark energy fluid.
Using the temporal behaviour  of dark energy density
as obtained from our analysis we 
obtain temporal behaviour of the $k$-essence scalar field $\phi$
for different values of the diffusion coefficient.
For any value of the diffusion coefficient within its allowed
range, the obtained time ($t$)-dependence of the scalar field
may be expressed as $\phi(t/t_0) = \phi_0 + \varepsilon_1 (t/t_0 - 1) + \varepsilon_2 (t/t_0 -1)^2$,
where $t_0$ is the time corresponding to present epoch.
The coefficients $\varepsilon_1$ and $\varepsilon_2$ are dependent on the
value of diffusion parameter. 
Interestingly the observational data do not allow further higher
order terms of $t$ in the obtained time dependence of field $\phi$
for any value of diffusion coefficient. 
Thus the $k-$essence scalar field $\phi$, in the context of this model,  
is found to have similar time evolution
properties of quintessence scalar field that drives homogeneous inflation.
Similar results in a different context were also obtained in \cite{arkapaper1}.\\

In Sec.\ \ref{sec:framework} we have discussed the basic frame work of
the model of  diffusive dark matter - dark energy interaction.
In Sec.\ \ref{sec:boundsdiffusion} we presented a brief description
of the methodology of analysis of SNe Ia data and describe how time dependences 
of some relevant cosmological
parameters extracted from the data has been exploited in the context of
the model to obtain bounds on diffusion parameter.
In Sec.\ \ref{sec:kessence} 
we have presented how we extracted the time evolution
of the $k-$essence scalar field $\phi$ from the observational
data, for a constant potential
in $k-$essence Lagrangian. We summarised the results obtained
in this work in Sec.\ \ref{sec:conclusion}.

\section{Model of interacting diffusive dark energy and dark matter}
\label{sec:framework}
We indulge in a brief outline of the model of diffusive dark matter - dark energy interaction.
The dynamics of cosmic evolution is governed by Einstein's equation
\begin{eqnarray}
R_{\mu\nu} - \frac{1}{2}g_{\mu\nu}R =  8\pi G T_{\mu\nu}
\label{eq:a1}
\end{eqnarray}
where $g_{\mu\nu}$ is the spacetime metric, $R_{\mu\nu}$
is the Ricci tensor, $R=g^{\mu\nu}R_{\mu\nu}$ and $G$
is Newton's gravitation constant. 
The total energy-momentum tensor, $T_{\mu\nu}$, 
of the universe, is conserved: $\nabla^\mu T_{\mu\nu} = 0$.
We write $T_{\mu\nu}$ by decomposing it into 
contributions from the different constituent 
components
(radiation($R$), baryonic matter($b$), dark matter(dm) and dark
energy(de)) of the universe as
\begin{eqnarray}
T_{\mu\nu} 
&=& T_{\mu\nu}^R +  T_{\mu\nu}^b +  T_{\mu\nu}^{\rm dm} +  T_{\mu\nu}^{\rm de}
\label{eq:a2}
\end{eqnarray}
When the dark sector (dark matter and dark energy) of
the universe is not interacting with baryonic matter and radiation,
the conservation of total energy momentum tensor implies
$\nabla^\mu (T_{\mu\nu}^R + T_{\mu\nu}^b) = 0$ and
$\nabla^\mu (T_{\mu\nu}^{\rm dm} + T_{\mu\nu}^{\rm de})=0$.
Now for interacting dark matter and dark energy resulting in
exchange of energy between then, we have 
\begin{eqnarray}
\nabla_\mu T_{de}^{\mu\nu} = - \nabla_\mu T_{dm}^{\mu\nu} \equiv  - \sigma J^\nu\,.
\label{eq:a3}
\end{eqnarray}
The transfer of energy between 
dark matter and dark energy is assumed to be caused by a 
diffusion in an ideal fluid environment.
$\sigma(>0)$  denotes
the diffusion coefficient which is a measure of 
average energy transferred to the particles 
of the dark matter fluid per unit time . $J^\mu$ denotes the current
density of matter 
satisfying  the conservation law $\nabla_\mu J^\mu = 0$.
We write the current density as $J^\mu = nu^\mu$, where $n$
is the number density of the particles of the  dark matter fluid
and $u^\mu$ represents the four-velocity of the fluid.
In a  homogeneous and isotropic 
spacetime background described by  
(FRW) metric with scale factor
$a(t)$, taking $u^\mu=(1,0,0,0)$ for the comoving fluid,
the conservation law $\nabla_\mu J^\mu = \nabla_\mu (nu^\mu) = 0$
implies $n(t) a^{3}(t) = {\rm constant} = n_0 $.
Here $n_0$ denotes number density at present epoch and value of the 
scale factor at present epoch
normalised to unity. Considering dark matter fluid to be
an ideal fluid characterised by its energy density $\rho_{\rm dm}$
and pressure $p_{\rm dm}$, the energy momentum tensor $T_{dm}^{\mu\nu}$
is given by
\begin{eqnarray}
T_{dm}^{\mu\nu} &=& \rho_{\rm dm} u^\mu u^\nu + p_{\rm dm}(g^{\mu\nu}+u^\mu u^\nu)
\label{eq:a4}
\end{eqnarray}
Using Eq.\  \eqref{eq:a4} and $J^\nu = nu^\nu$  in Eq.\ \eqref{eq:a3}
and projecting it along the direction of $u^\nu$ we obtain
\begin{eqnarray}
 \dot{\rho}_{\rm dm} + 3H\rho_{\rm dm}  =   \sigma \frac{n_0}{a(t)^3}
 \label{eq:a5}
\end{eqnarray}
where $H = \dot{a}/a$ is the Hubble parameter and 
we take dark matter as pressureless dust ($p_{\rm dm}=0$).
We also model the dark energy as an ideal fluid 
characterised by  its energy density ($\rho_{\rm de}$)
and pressure ($p_{\rm de}$). In FRW background spacetime,
conservation of total energy momentum tensor 
for the dark fluid, $\nabla^\mu (T_{\mu\nu}^{\rm dm} + T_{\mu\nu}^{\rm de})=0$,  
implies
\begin{eqnarray}
(\dot{\rho}_{\rm dm} + \dot{\rho}_{\rm de}) + 3 H \Big{[}(\rho_{\rm dm} + \rho_{\rm de}) + p_{\rm de}\Big{]} &=& 0\,.
\label{eq:a6}
\end{eqnarray}
From Eq.\  \eqref{eq:a5} and \eqref{eq:a6} we have
\begin{eqnarray}
\dot{\rho}_{\rm de}+3H(\rho_{\rm de}+p_{\rm de}) 
&=&  -\sigma \frac{n_0}{a(t)^3}
\label{eq:a7}
\end{eqnarray}

\section{Bounds on the diffusion coefficient from SNe Ia data}
\label{sec:boundsdiffusion}
In this section we first discuss methodology to 
extract the temporal behaviour of the
FRW scale factor $a(t)$, energy density and pressure of 
the dark fluid
during the late time phase of cosmic evolution
from the analysis of SNe Ia data. Then we discuss
how we exploit the obtained time dependence of scale factor
to find bounds on the diffusion coefficient term $\sigma n_0$ appearing
in RHS of Eqs.\  \eqref{eq:a5} and \eqref{eq:a7}. \\

The  luminosity distance and  redshift relationship
for redshift values up to $z \sim 1$ obtained from SNe Ia observations
is instrumental in revealing features of late time phase
of cosmic evolution. At
present there exists several dedicated and systematic 
searches and measurements of SNe Ia. These include
supernova surveys in different redshift domains.
The high redshift ($z \sim 1$) projects are Supernova Legacy Survey (SNLS) 
(\cite{ref:Astier},\cite{ref:Sullivan}) the ESSENCE project \cite{ref:Wood-Vasey}, 
the Pan-STARRS survey (\cite{ref:Tonry},\cite{ref:Scolnic},\cite{ref:Rest})
Searches in redshift regime $0.05<z<0.4$ are performed in 
The SDSS-II supernova surveys (\cite{ref:Frieman},\cite{ref:Kessler},\cite{ref:Sollerman},
\cite{ref:Lampeitl},\cite{ref:Campbell}).
Small redshift programmes $(z > 0.1)$
include the Harvard-Smithsonian Center for Astrophysics survey (cFa) \cite{ref:Hicken}, 
the Carnegie Supernova Project (CSP)(\cite{ref:Contreras},\cite{ref:Folatelli},\cite{ref:Stritzinger}) 
the Lick Observatory Supernova Search (LOSS) 
\cite{ref:Ganeshalingam}  and the Nearby Supernova Factory (SNF) \cite{ref:Aldering}. 
In all these surveys around one thousand SNe IA events 
were discovered. The measured   luminosity distance 
has a very high statistical precision in the range between
$z \sim 0.01$ and $z \sim 0.7$.
Recently, ``Joint Light-curve Analysis" (JLA) data 
(\cite{ref:Scolnic},\cite{ref:Conley},\cite{ref:Suzuki}) has been released.
The data consists of 740 SNe Ia events.
This includes a new compilation of SNe Ia light curves
including data from the full three years of the SDSS survey, 
first three seasons of the five-year SNLS survey and  14 very
high redshift $0.7 < z < 1.4$ SNe Ia from space-based observations 
with the HST \cite{ref:Riess}. 
This data sample has been extensively studied analysed in 
recent years. 
The data sets provide observed values
of distance modulus at different measured values of red-shift ($z$)
We use the results of analysis of JLA data following flux-averaging technique described in
\cite{ref:wangapj,wang:jcap,wang:prd} which takes care of different systematic uncertainties 
of SNe IA data in a elegant way.
The $\chi^2$ function of JLA data is given by 
\begin{eqnarray*}
\chi^2 = \sum_{i,j}(\mu_{\rm obs}  - \mu_{\rm th})_i
(\sigma^{-1})_{ij}
(\mu_{\rm obs}  - \mu_{\rm th} )_j
\label{eq:b1}
\end{eqnarray*}
where $\mu_{\rm obs}$ is the observed value of 
distance modulus at red-shift $z_i$. Theoretically,
the distance modulus is related to the luminosity
distance $d_L$ by $\mu_{\rm th} = 5\log_{10} [d_L/{\rm Mpc}] + 25$ and 
 is expressed as $\mu = m_B^\star - (M_B - \alpha \times X_1 + \beta\times C)$
where $m_B^\star$ is the observed peak magnitude, $\alpha$, $\beta$ and $M_B$
are nuisance parameters. $X_1$ is the time
stretching of the light-curve and $C$ 
describes the supernova color at maximum brightness as defined in detail in \cite{jla}.
$\sigma$ is the covariant matrix involving
statistical and systematic uncertainties  
as defined by Eq.\ (2.16) of \cite{wang:jcap}.
For the details of calculation of 
$\chi^2$ we refer the reader to Refs.\ \cite{wang:jcap} and \cite{wang:prd}
where a comprehensive analysis of JLA data
has been performed.
A red-shift cut-off ($z_{\rm cut}$) is used there
to separate out SN samples with $z < z_{\rm cut}$
and $z \geq z_{\rm cut}$. For samples with $z < z_{\rm cut}$
the $\chi^2$ has been computed using Eq.\ (\ref{eq:b1})
and for samples with $z \geq z_{\rm cut}$ flux averaged
values of $\mu$ and covariant matrix are used to
compute $\chi^2$  in a way described in detail in \cite{wang:jcap}.
For our work we take as input, the $z-$dependence of the function $E(z) = H(z)/H_0$,
obtained from the marginalisation of $\chi^2$ over parameters
$\alpha$, $\beta$, $M_B$  etc. as
shown in left panel 
of Fig.\ 5 of Ref.\ \cite{wang:jcap}. We take $E(z)$ vs $z$ curve
obtained in  \cite{wang:jcap} for two benchmark cases: 
$z_{\rm cut}=0$ and $z_{\rm cut}=0.6$.\\

We may use the relations $H=\dot{a}/a$ and 
$a_0/a = 1+z$ to write
\begin{eqnarray}
dt &=& - \frac{dz}{(1+z)H(z)} = - \frac{dz}{(1+z)H_0E(z)}
\label{eq:b2}
\end{eqnarray}
The above equation on integration gives
\begin{eqnarray}
\frac{t(z)}{t_0} &=& 1 
- \frac{1}{H_0t_0}\int_z^0 \frac{dz^\prime}{(1+z^\prime)E(z^\prime)}
\label{eq:b3}
\end{eqnarray}
where $t_0$ is the time denoting the present epoch. 
The function $E(z)$ as obtained from analysis of JLA data in \cite{wang:jcap}
is used in Eq.\ \eqref{eq:b3}, to obtain $t$ as a
function of $z$ by performing the integration numerically.
We then eliminate $z$ from the obtained $z$ - $t(z)$ dependence and the
the equation $a_0 /a = 1 + z$ to obtain scale factor $a$ as a function
of $t$. 

In FRW spacetime background, 
the equations governing dynamics of late time cosmic evolution are the following two independent Friedmann equations
\begin{eqnarray}
&& H^2 = \frac{8\pi G}{3} (\rho_{\rm de} + \rho_{dm}) \label{eq:bb1}\\
&& \frac{\ddot{a}}{a} = - \frac{4\pi G}{3}\left[(\rho_{\rm dm} + \rho_{\rm de}) + 3p_{\rm de} \right] \label{eq:bb2}
\end{eqnarray}
Equation of state of total dark fluid may then
be expressed in terms of scale factor and its 
derivatives from above two equations as
\begin{eqnarray}
\omega 
= \frac{p_{\rm de}}{\rho_{\rm de} + \rho_{\rm dm}}
= -\frac{2}{3}\frac{\ddot{a}a}{\dot{a}^2} - \frac{1}{3}
\label{eq:bb3}
\end{eqnarray}
We have considered a flat spacetime (zero curvature constant)
and neglect contributions from radiation and baryonic matter during late time phase of cosmic evolution.\\

\begin{figure}[t]
\begin{center}
\includegraphics[scale=.55]{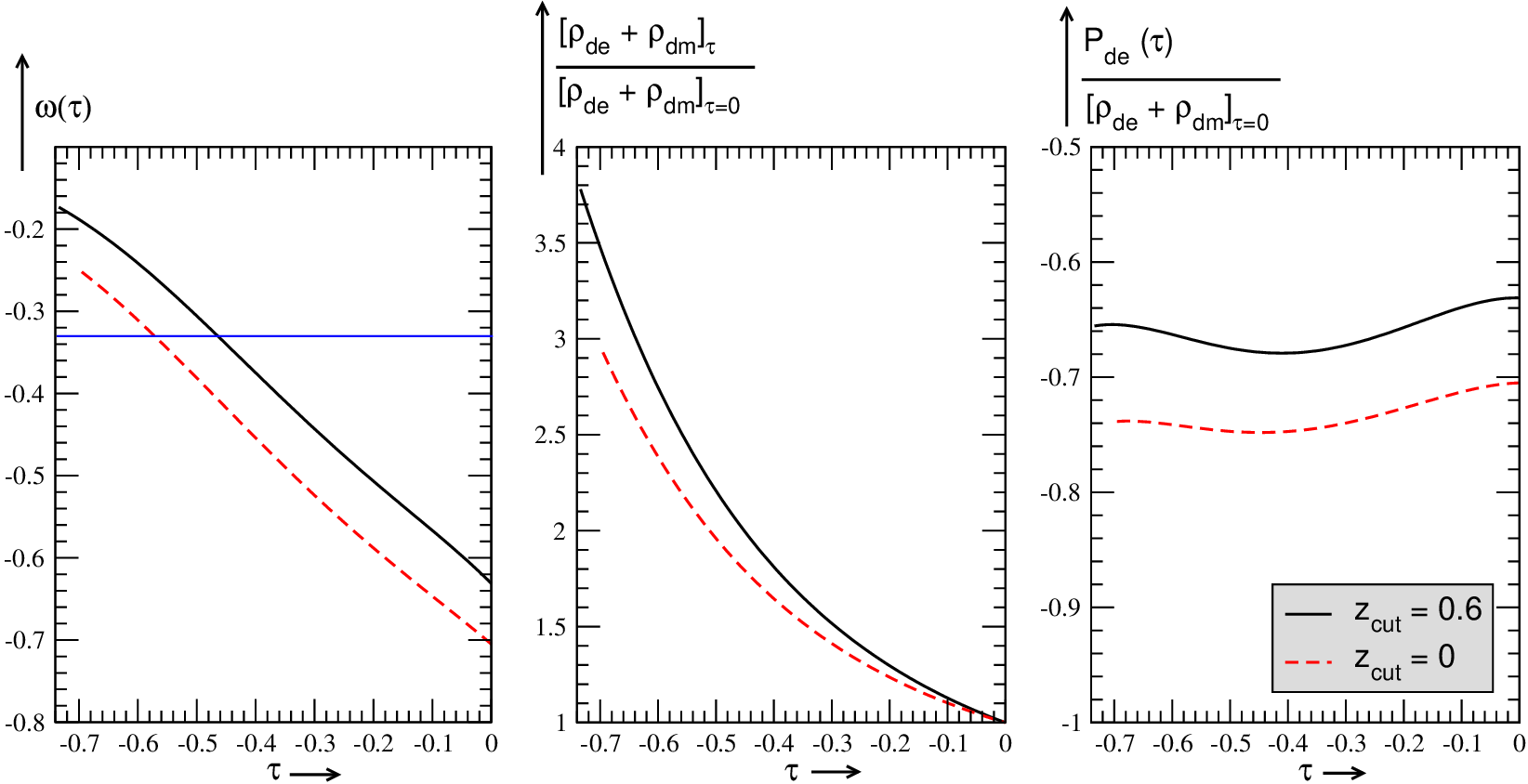} 
\end{center}
\caption{\label{fig:1}
Left panel: Plot of $\omega(\tau)$ vs $\tau$ as obtained from analysis of observational data
for $z_{\rm cut} = 0$ and $z_{\rm cut} = 0.6$  The horizontal line represent the value $\omega=-1/3$ ($\ddot{a}$ =0),
Middle panel: Plot of $\frac{[\rho_{\rm de} + \rho_{\rm dm} ]_\tau
}{ [\rho_{\rm de} + \rho_{\rm dm} ]_0}$ vs $\tau$ as obtained from analysis of observational data
for $z_{\rm cut} = 0$ and $z_{\rm cut} = 0.6$,
Right panel: Behaviour of $p_{de}$ vs $\tau$ for $Z_{\rm cut}$=0 and $Z_{\rm cut}$=0.6 }
\end{figure}

The time dependence of equation of state $\omega$ 
of the dark fluid may be obtained by
using time dependence of scale factor  
in Eq.\  \eqref{eq:bb3}. We express the temporal behaviour
in terms of a dimensionless time parameter 
$\tau$ defined as 
\begin{eqnarray}
\tau = \ln a(t)
\label{eq:bb4}
\end{eqnarray}
The time domain accessible in 
Supernova Ia observations is  $-0.7 < \tau <0$. 
$\tau=0$ corresponds to present epoch as
the value of scale factor at present
epoch is normalised to unity.
 In Fig.\ \ref{fig:1} we have shown time($\tau$)-dependence
 of equation of state $\omega$.   
We find that the obtained dependence as shown in Fig.\ \ref{fig:1} 
may be fitted with a 
polynomial of the form
\begin{eqnarray}
\omega(\tau) = -1 + \sum_{i=0} B_i \tau^i
\label{eq:bb5}
\end{eqnarray}
with coefficients $B_i$'s given in Tab.\ \ref{tab:1}.
\begin{table}[h]
\begin{center}
\begin{tabular}{|cc|cc|cc|}
\hline
$B_0 = $ & -0.704 (-0.631) & $B_3 =$ & -2.29 (-3.76) &&\\
\cline{1-4}
$B_1 = $ & -0.61 (-0.715)  & $B_4 =$ & -2.81 (-4.84) & $B_i = 0$&\\ 
\cline{1-4}
$B_2 = $ & -0.49 (-1.04)  & $B_5 =$ & -0.92 (-1.93) & for $i>5$ &\\
\hline
\end{tabular}
\end{center}
\caption{\label{tab:1} Values of $B_i$'s in Eq.\  \eqref{eq:bb5}
providing best fit to the values of $\omega(\tau)$
extracted from SNe Ia data samples corresponding to $z_{\rm cut}=0$ ($z_{\rm cut}=0.6$).}
\end{table}

In terms of the newly defined time parameter $\tau$,
the continuity equation   \eqref{eq:a6}  for the 
total dark fluid may be written as
\begin{eqnarray}
\done{}{\tau} \ln\Big{(} \rho_{\rm de} + \rho_{\rm dm}\Big{)}
&=& -3\Big{(}1 +\omega(\tau)\Big{)}\,,
\label{eq:bb6}
\end{eqnarray}
which on integration gives
\begin{eqnarray}
\Big{[}\rho_{\rm de} + \rho_{\rm dm} \Big{]}_\tau
&=&
{\Big{[}\rho_{\rm de} + \rho_{\rm dm} \Big{]}_0}
\exp\left[-3\int^{\tau}_{\tau^\prime=0}(1+\omega(\tau^\prime)) 
d\tau^\prime \right] 
\label{eq:bb7}
\end{eqnarray}
Using the   temporal dependence of the function 
$\omega(\tau)$ as obtained and depicted in Fig.\ \ref{fig:1},
we perform the integration 
appearing on the right hand side of Eq.\ \eqref{eq:bb7}
numerically to obtain $\tau-$dependence  of the 
total energy density of the dark fluid. The obtained
temporal behaviour is shown in middle panel of Fig.\ \ref{fig:1}.
We find that this time dependence   may be
expressed in terms of a fitted polynomial of the form
\begin{eqnarray}
 \Big{[}\rho_{\rm de} + \rho_{\rm dm}\Big{]}_{\tau} &=& \big{[}\rho_{\rm de} + \rho_{\rm dm}\big{]}_0 \sum_{i=0} C_i \tau^i
\label{eq:bb8}
\end{eqnarray}
with coefficients ($C_i$'s) given in Tab.\ \ref{tab:2}.\\
\begin{table}[t]
\begin{center}
\begin{tabular}{|cc|cc|cc|}
\hline
$C_0 = $ & 1 & $C_3 =$ &  -0.65 (-0.62)  && \\
\cline{1-4}
$C_1 = $ & -0.89 (-1.10) & $C_4 =$ & 1.36 (2.096) & $C_i=0$ &\\ 
\cline{1-4}
$C_2 = $ & 1.28 (1.65)  & $C_5 =$ & -0.97 (-1.05) & for $i>5$ &\\
\hline
\end{tabular}
\end{center}
\caption{\label{tab:2}Values of $C_i$'s in Eq.\  \eqref{eq:bb8} 
providing best fit to the values of ($\rho_{\rm de}(\tau) + \rho_{\rm dm}(\tau)/ 
[\rho_{\rm de} + \rho_{\rm dm}]_0$) extracted from SNe Ia data samples corresponding to $z_{\rm cut}=0$ ($z_{\rm cut}=0.6$).}
\end{table}

Using the obtained time dependences of $\omega$ and
$\rho_{\rm de} + \rho_{\rm de}$ in Eq.\ \eqref{eq:bb3},
\begin{eqnarray}
p_{\rm de}(\tau) &=& \omega(\tau) \Big{[}\rho_{\rm de} + \rho_{\rm de}\Big{]}_{\tau}
\label{eq:bb9}
\end{eqnarray}
we may also obtain the temporal behaviour of pressure of
the dark energy fluid. The obtained dependence is shown in right panel of  Fig.\ \ref{fig:1}
and  may be expressed in terms
of a fitted polynomial 
\begin{eqnarray}
p_{\rm de}(\tau) &=& \big{[}\rho_{\rm de} + \rho_{\rm dm}\big{]}_0 \sum_{i=0} \gamma_i \tau^i
\label{eq:bb10}
\end{eqnarray}
with coefficients ($\gamma_i$'s) given in Tab.\ \ref{tab:3}.\\
\begin{table}[t]
\begin{center}
\begin{tabular}{|cc|cc|cc|}
\hline
$\gamma_0 = $ & -0.705 (-0.631) & $\gamma_3 =$ & -2.27 (-3.463)   & $\gamma_i=0$ & \\
\cline{1-4}
$\gamma_1 = $ & 0.0016 (-0.0319) & $\gamma_4 =$ &  -1.48 (-2.288) &  for $i>4$ &\\ 
\cline{1-2}
$\gamma_2 = $ &  -0.93 (-1.395) &   &   &  &\\
\hline
\end{tabular}
\end{center}
\caption{\label{tab:3}Values of $\gamma_i$'s  in Eq.\  \eqref{eq:bb10} 
providing best fit to the values of   ($p_{\rm de}(\tau) / 
[\rho_{\rm de} + \rho_{\rm dm}]_0$) extracted from  SNe Ia data samples corresponding to $z_{\rm cut}=0$ ($z_{\rm cut}=0.6$). }
\end{table}

Time evolution of the dark matter energy density $\rho_{\rm dm}$ in this model is
given by Eq.\  \eqref{eq:a5}. In terms of  time parameter   $\tau$
the equation may be rewritten as
\begin{eqnarray}
\frac{d\rho_{\rm dm}}{d\tau} + 3\rho_{\rm dm}
&=&  \frac{\sigma n_0}{a^3(\tau) H(\tau)}  
\label{eq:c1}
\end{eqnarray}
Using $\tau$-dependence of the quantity $\frac{1}{a^3H}$ appearing
in right hand side of above equation may be known using obtained
temporal behaviour of the scale factor.
We find that this dependence may be expressed in terms of a fitted
polynomial 
\begin{eqnarray}
\frac{1}{a^3(\tau) H(\tau)} & = &  \sum_{i=0}^5 D_{i}\tau^i 
\label{eq:c2}
\end{eqnarray}
with coefficients ($D_i$'s) given in Tab.\ \ref{tab:4}. 
\begin{table}[]
\begin{center}
\begin{tabular}{|cc|cc|c|}
\hline
$D_0 = $ &  1    &   $D_3 =$  & -2.70 &$D_i = 0$\\
\cline{1-4}
$D_1 = $ & -2.56  & $D_4 =$  & -1.83 & for $i>5$ \\ 
\cline{1-4}
$D_2 = $ &  2.65  & $D_5 =$  &  -0.936 &\\
\hline
\end{tabular}
\end{center}
\caption{\label{tab:4} Values of  $D_i$'s in Eq.\  \eqref{eq:c2} 
providing best fit to the values of $\frac{1}{a^3(\tau) H(\tau)}$
extracted from  SNe Ia data samples corresponding to $z_{\rm cut}=0$}
\end{table}
We now assume a series solution of Eq.\  \eqref{eq:c1} for  $\rho_{\rm dm}$
as
\begin{eqnarray}
\rho_{\rm dm} & = & \Big{[}\rho_{\rm de} + \rho_{\rm dm} \Big{]}_0 \sum_{i=0}^\infty\alpha_{i}\tau^i 
\label{eq:c3}
\end{eqnarray}
Substituting Eqs.\  \eqref{eq:c2} and \eqref{eq:c3} in Eq.\  \eqref{eq:c1}
we obtain
\begin{eqnarray}
\sum_{i=0}^\infty i\alpha_{i}\tau^{i-1} + 3 \sum_{i=0}^\infty\alpha_{i}\tau^i = K \sum_{i=0}^5 D_{i}\tau^i 
\label{eq:c4}
\end{eqnarray}
where
\begin{eqnarray}
K &=& \frac{\sigma n_0}{[\rho_{\rm de} + \rho_{\rm dm} ]_0}
\label{eq:c5}
\end{eqnarray}
Equating the coefficients of $\tau^{i}$ from both sides of Eq.\  \eqref{eq:c4} we obtain
\begin{eqnarray}
\alpha_{i+1} &=& \frac{K  D_i - 3\alpha_i}{i+1} 
\label{eq:c6}
\end{eqnarray}
Note that  $K(>0)$ represents a diffusion parameter as it  is   linearly related to  
diffusion coefficient $\sigma$. We also note
from Eq.\  \eqref{eq:c3} that $\alpha_0$ corresponds
to the value of the fraction $\frac{\rho_{\rm dm}}{[\rho_{\rm de} + \rho_{\rm dm} ]_0}$
at $\tau = 0$ (present epoch). 
Mathematically, $\alpha_0$ is thus defined in the domain 
$0< \alpha <1$. Now for a given set of values for $\alpha_0$ and $K$,
one may find $\alpha_i$'s ($i>0$) using the recursion relation  \eqref{eq:c6}.
Since $D_i$'s are zero for $i>5$ (see Tab.\ \ref{tab:4}) and
the term $(i+1)$ appears in the denominator of the recursion relation,
the evaluated series  \{$\alpha_i$\} will always be convergent.
Using these values
of $\alpha_i$'s   we may compute
$\rho_{\rm dm}$ at all values of $\tau$ from Eq.\ \eqref{eq:c3}. Since 
$|\tau|<1$, and the series  \{$\alpha_i$\} is convergent, 
evaluated value of $\rho_{\rm dm}$ gets negligible contribution
from terms above certain order in the summation series in Eq.\ \eqref{eq:c3}. 
\footnote{For example we find that values of $\rho_{\rm dm}$ computed with first
6 terms of the series and with first 100 terms of the series differ by less than 1 percent.}\\

\begin{figure}[!ht]
\begin{center}
\includegraphics[scale=.4]{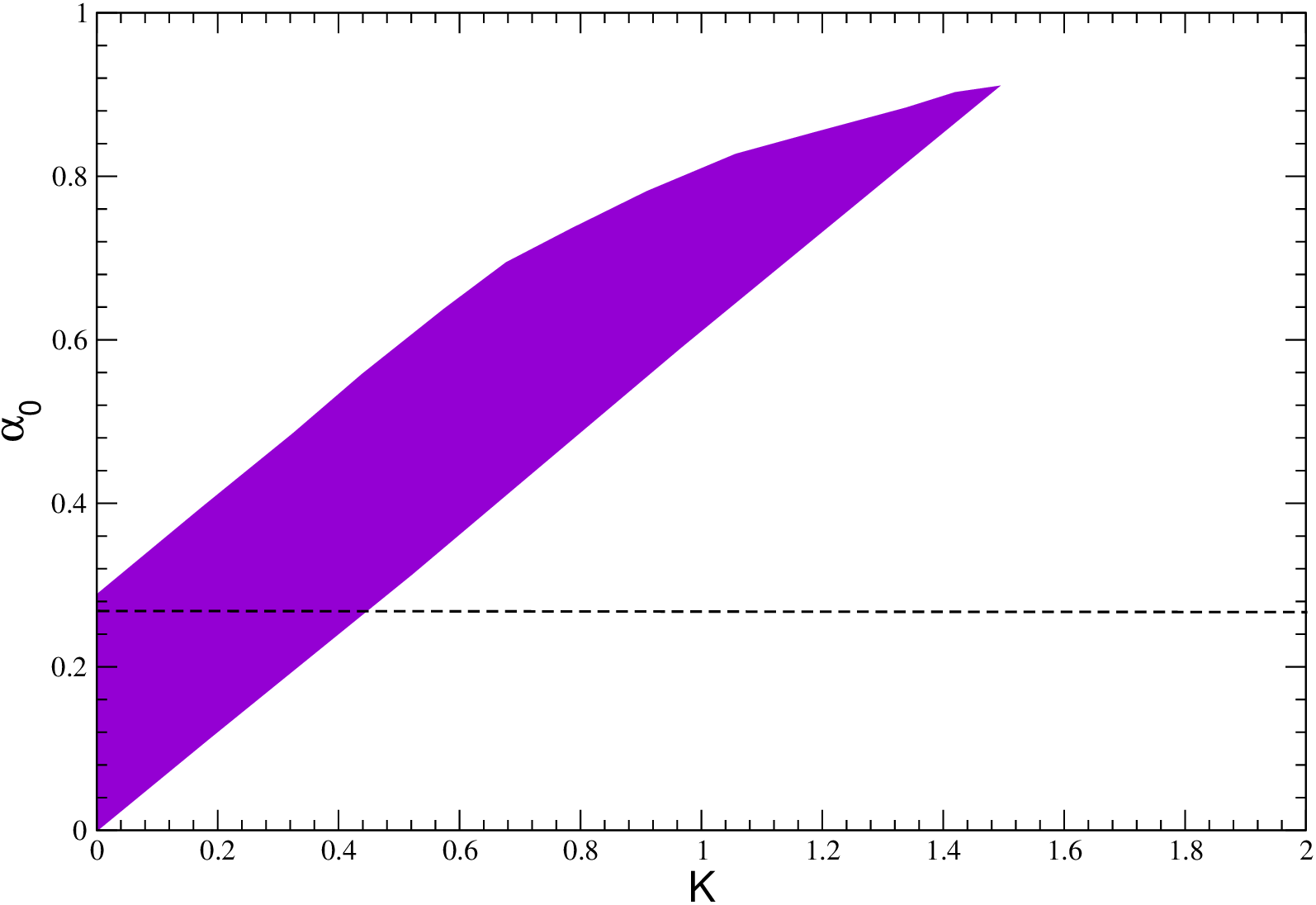} 
\end{center}
\caption{\label{fig:2} Region of  $\alpha_0 - K$ parameter space}
\end{figure}

The values of energy density  $\Big{[}\rho_{\rm dm} + \rho_{\rm de}\Big{]}_\tau$ 
of the total dark fluid at any instant of time $\tau$ has  
been obtained directly from the analysis of SNe Ia data and shown in 
Fig.\ \ref{fig:1} (middle panel). The value of dark matter density $\rho_{\rm dm}(\tau; \alpha_0,K)$
computed from Eq.\ \eqref{eq:c3} at any $\tau$,
for a given $(\alpha_0,K)$ is subject to the constraint 
\begin{eqnarray}
0 < \rho_{\rm dm}(\tau; \alpha_0,K) <\Big{[}\rho_{\rm dm} + \rho_{\rm de}\Big{]}_\tau
\label{eq:c7}
\end{eqnarray}
for  all values of $\tau$ in the range
($-0.7 < \tau<0$) accessible in SNe Ia observations. Imposition of the
constraint (Eq.\ \eqref{eq:c7}) limits the range of allowed values of $\alpha_0$ and $K$.
The shaded region in Fig.\ \ref{fig:2} depicts the allowed domain
in $\alpha_0 - K$ parameter space for which the constraint in Eq.\ \eqref{eq:c7}
is realised.\\

However, measurements in satellite borne experiments
- WMAP \cite{Hinshaw:2012aka} and Planck \cite{Ade:2013zuv}  
established that fractional contribution of dark matter 
to the total energy density
of present universe is  $\sim 0.27$.
So neglecting contributions from radiation and baryonic matter
to the total energy density of present universe,
we may take measured value of $\alpha_0$ to be $\sim 0.27$.
This has been depicted by a horizontal line in Fig.\ \ref{fig:2}.
This value of $\alpha_0$ corresponds to an allowed range of
diffusion parameter $K$ as $0 \leqslant K < 0.44$.

\section{Diffusive dark energy - dark matter scenario with dark energy as a $k$-essence scalar field}
\label{sec:kessence}
In this section we realise dark energy in terms of a $k-$essence scalar field 
and investigate its implications in the context of
diffusive   dark matter - dark energy  model.
We assume dark energy to be represented by a homogeneous scalar field $\phi$ 
whose dynamics is driven by a $k$-essence Lagrangian with constant potential.
Using the temporal behaviour of scale factor, total energy density of dark fluid,
pressure of the dark energy fluid, as extracted from SNe Ia data, we find
nature of time dependence of the scalar field $\phi$. 
These has been obtained for different values of the diffusion coefficient, $K$
within it's range allowed from SNe Ia data as obtained in Sec.\ \ref{sec:boundsdiffusion}.
The dependence of the field on diffusion coefficient has also been obtained. \\

\begin{figure}[!t]
\begin{center}
\includegraphics[scale=.4]{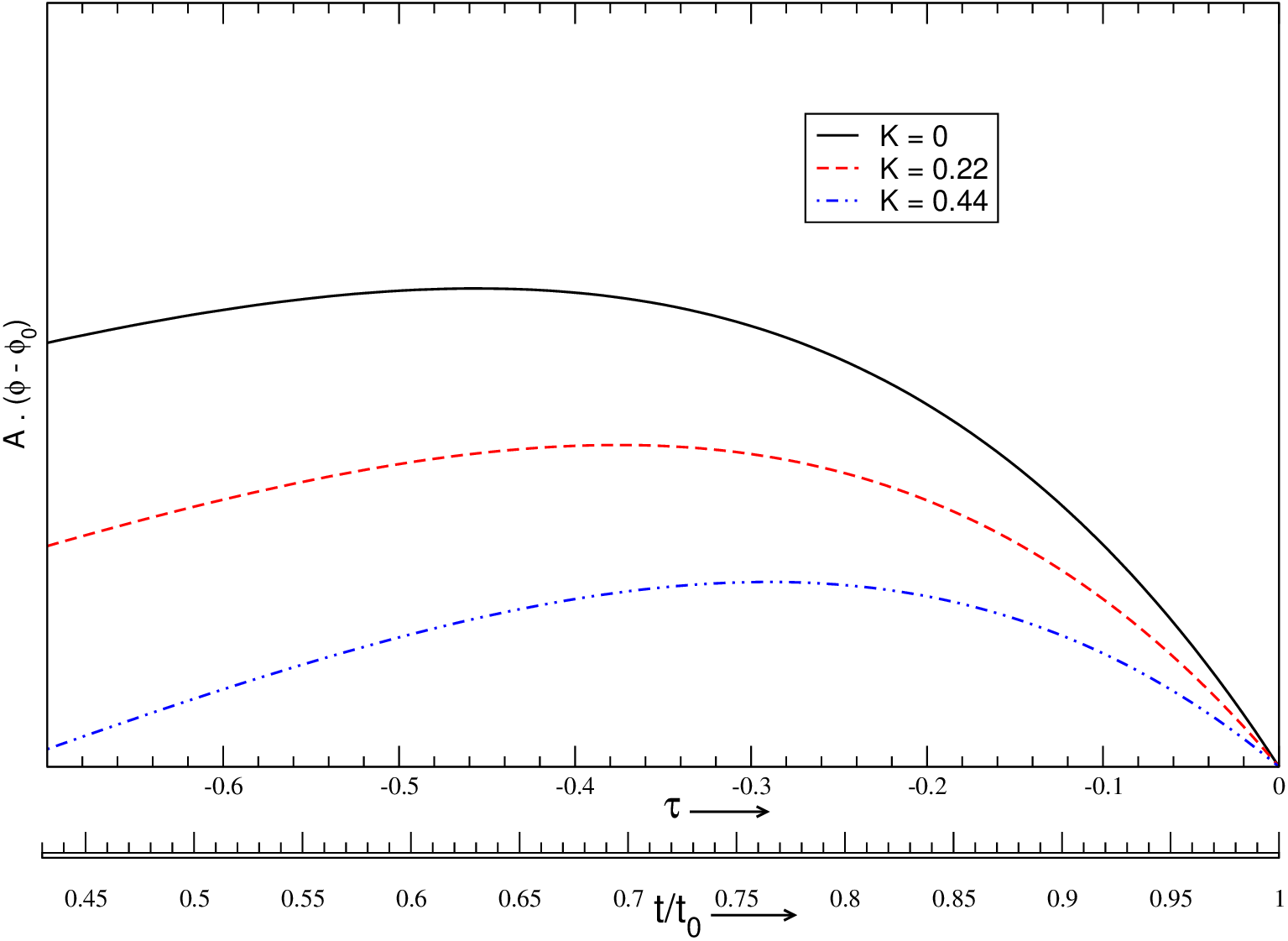} 
\end{center}
\caption{\label{fig:4} Temporal behaviour of the quantity $A (\phi -\phi_0)$,
(where  $A \equiv   \sqrt{2C}V \Big{/}(\rho_{\rm dm}^0 + \rho_{\rm de}^0) $
is a constant)
for three benchmark  values of diffusion parameter: $K = 0, 0.22, 0.44$. }
\end{figure}

The $k-$essence models involve non-canonical Lagrangian of the form
$L = V(\phi)F(X)$ where $X = (1/2)g^{\mu\nu}\nabla_\mu\phi\nabla_\nu\phi$,
$F$ and $V$ are  functions of $X$ and $\phi$ respectively. The stress energy
tensor corresponding to this Lagrangian is equivalent to that of an 
ideal fluid with energy density $V(\phi)(2XF_X - F)$ and pressure $V(\phi)F(X)$ respectively,
where $F_X = dF/dX$. Identifying this fluid as dark energy we write 
\begin{eqnarray}
p_{\rm de} &=&  V   F(X) \,, \label{eq:dk1}\\
\rho _{\rm de} &=& V (2XF_X-F)\, .\label{eq:dk2}
\end{eqnarray}
Here we consider the $k-$essence model with constant potential,
$V(\phi) = V$ and the scalar field $\phi$  in FRW spacetime background 
to be homogeneous: $\phi(x) \equiv \phi(t)$. We then have
$X = (1/2)\dot{\phi}^2$ and 
constancy of the potential $V$ ensures existence of 
scaling relation \cite{scale1,scale2}
\begin{eqnarray}
XF_X^2 &=& C a^{-6}\,, \quad \mbox{$C$ is a constant}
\label{eq:dk3}
\end{eqnarray}
From Eqs.\ \eqref{eq:dk1} and \eqref{eq:dk2} we have
\begin{eqnarray}
\rho _{\rm de} + p_{\rm de} 
&=& 2VXF_X \label{eq:dk4}
\end{eqnarray}
Eliminating $X$ from Eqs.\ \eqref{eq:dk3} and \eqref{eq:dk4} we obtain
\begin{eqnarray}
\frac{F_X}{2CV} &=& \frac{a^{-6}}{\rho _{\rm de} + p_{\rm de}}
\label{eq:dk5}
\end{eqnarray}
Eliminating $F_X$ from Eq.\  \eqref{eq:dk3} and Eq.\  \eqref{eq:dk5} we obtain
\begin{eqnarray}
X &=& \frac{a^6 (\rho_{\rm de} + p_{\rm de})^2}{4CV^2} 
\label{eq:dk6}
\end{eqnarray}
For a homogeneous $k-$essence field $\phi$, we have $X = \frac{1}{2}\dot{\phi^2}$. Changing
the time parameter from $t$ to $\tau$ we may write
\begin{eqnarray}
X &=&  \frac{1}{2} \left[H  \left(\frac{d\phi}{d\tau}\right)\right]^2
\label{eq:dk7}
\end{eqnarray}
From Eq.\  \eqref{eq:dk6} and  \eqref{eq:dk7} we have
\begin{eqnarray} 
\left[\frac{\sqrt{2C}V}{(\rho_{\rm dm}^0 + \rho_{\rm de}^0)} \right] \left(\frac{d\phi}{d\tau}\right)  
&=&
\frac{a^3}{H} 
\left[\frac{\rho_{\rm de}}{(\rho_{\rm dm}^0 + \rho_{\rm de}^0)} + \frac{p_{\rm de}}{(\rho_{\rm dm}^0 + \rho_{\rm de}^0)}\right] 
\label{eq:dk8}
\end{eqnarray}
Eq.\  \eqref{eq:dk8} on Integration gives
\begin{eqnarray}
\left[\frac{\sqrt{2C}V}{(\rho_{\rm dm}^0 + \rho_{\rm de}^0)} \right] 
(\phi(\tau, K) - \phi_0(K))
&=&
\int_{\tau^\prime=0}^\tau d\tau^\prime \left[\frac{a^3(\tau^\prime)}{H(\tau^\prime)} 
\left(\frac{\rho_{\rm de}(\tau^\prime)}{\rho_{\rm dm}^0 + \rho_{\rm de}^0} 
+ \frac{p_{\rm de}(\tau^\prime)}{\rho_{\rm dm}^0 + \rho_{\rm de}^0}\right)  \right]
\label{eq:dk9}
\end{eqnarray}
The temporal behaviour of the quantities
$\Big{[}\rho_{\rm dm} + \rho_{\rm de}\Big{]}_\tau \Big{/}(\rho_{\rm dm}^0 + \rho_{\rm de}^0) $  and   
 $p_{\rm de}(\tau) \Big{/}(\rho_{\rm dm}^0 + \rho_{\rm de}^0) $
extracted from SNe Ia observations have been shown in  Fig.\ \ref{fig:1}. 
As discussed in Sec.\ \ref{sec:boundsdiffusion},
using Eq.\ \eqref{eq:c3} we may compute dark matter density 
$\rho_{\rm dm}(\tau; \alpha_0,K) \Big{/}(\rho_{\rm dm}^0 + \rho_{\rm de}^0) $
corresponding to a set of values of parameters ($\alpha_0,K$) within their
allowed domain depicted in Fig.\ \ref{fig:2}. 
The dark energy density $\rho_{\rm de}$ may
also be evaluated at a given  ($\alpha_0,K$) value as
\begin{eqnarray}
\frac{\rho_{\rm de}(\tau; \alpha_0,K)}{(\rho_{\rm dm}^0 + \rho_{\rm de}^0)} 
&=& \frac{\Big{[}\rho_{\rm dm} + \rho_{\rm de}\Big{]}_\tau}{(\rho_{\rm dm}^0 + \rho_{\rm de}^0)} 
- \frac{\rho_{\rm dm}(\tau; \alpha_0,K)}{(\rho_{\rm dm}^0 + \rho_{\rm de}^0)}  \label{eq:dk10}
\end{eqnarray}
\begin{figure}[!t]
\begin{center}
\includegraphics[scale=.4]{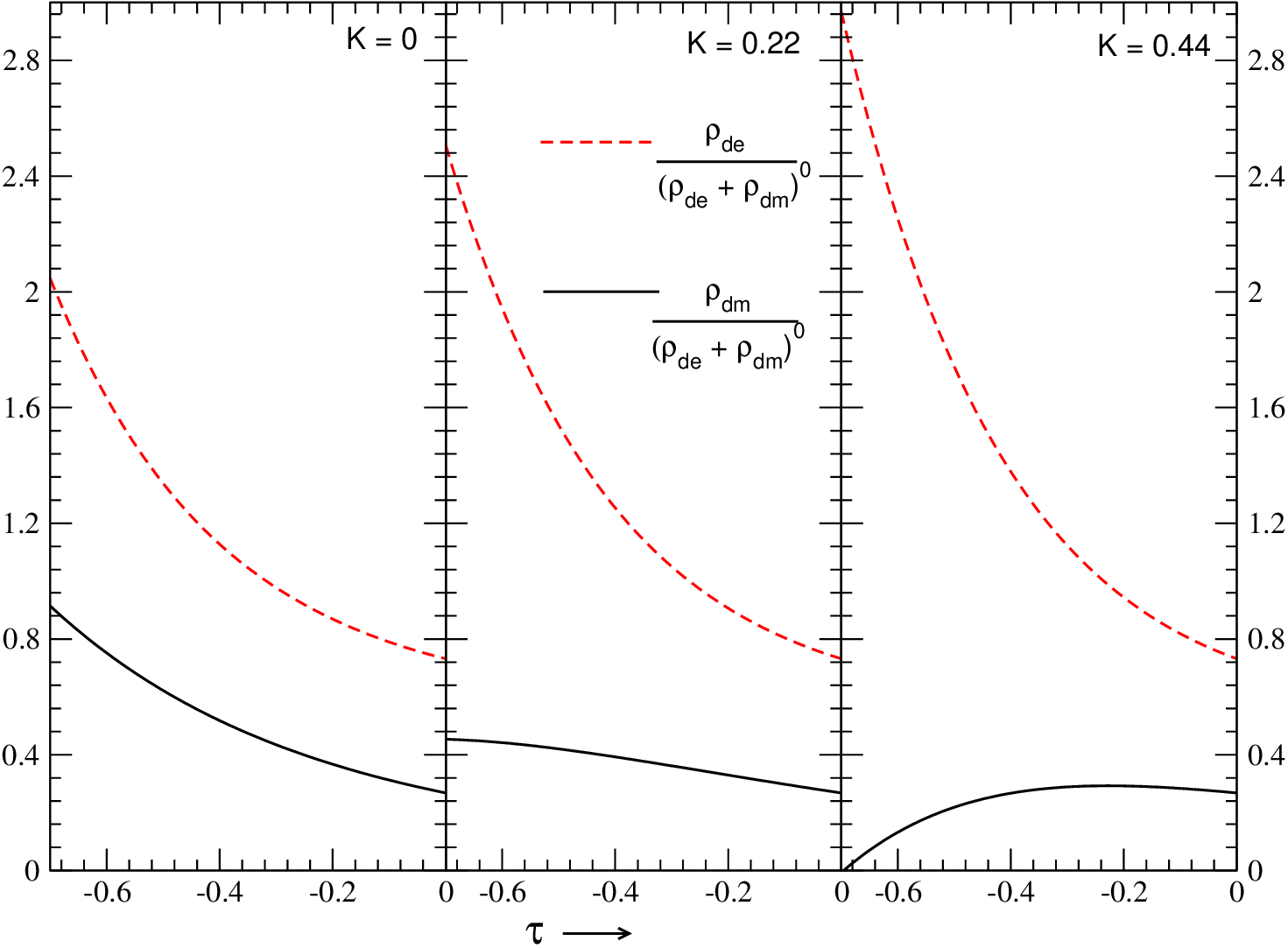} 
\end{center}
\caption{\label{fig:3} Temporal behaviour for dark matter density  
$\rho_{\rm dm}$ and 
dark energy density $\rho_{\rm de}$
for three benchmark  values of diffusion parameter: $K = 0, 0.22, 0.44$}
\end{figure}
We compute $\rho_{\rm de}$ from above equation for $\alpha_0 \equiv 
\rho_{\rm dm}^0\Big{/}(\rho_{\rm dm}^0 +\rho_{\rm de}^0) = 0.27$ 
(which is close to the experimentally observed value)
and for different values of $K$ in the corresponding allowed range $0< K < 0.44$.
The obtained temporal behaviour for dark matter density  $\rho_{\rm dm}$ and 
dark energy density $\rho_{\rm de}$
for three benchmark  values of diffusion parameter $K$ (0, 0.22 and 0.44) are shown
in Fig.\ \ref{fig:3}.
With the  obtained $\tau$ dependences of scale factor 
$a$, Hubble parameter $H$, $p_{\rm de}(\tau)$ and $\rho_{\rm de}(\tau, K)$, we compute 
values of integrand appearing in the right hand side of Eq.\  \eqref{eq:dk9}
at different values of $\tau$ and $K$. We then perform the integration numerically 
to obtain temporal behaviour of 
the $k$-essence scalar field for different values of the 
diffusion parameter $K$. The results are shown in Fig.\ \ref{fig:4}.
We have shown the time dependence in terms of both the time parameters $\tau = \ln a(t)$ and $t$.
We find that, for any value of the diffusion parameter $K$,
the time dependence of the $k-$essence scalar field $\phi$
may be fitted in terms of polynomial of degree 2 as
\begin{eqnarray}
\phi(t/t_0) = \phi_0 + \varepsilon_1 (t/t_0 - 1) + \varepsilon_2 (t/t_0 -1)^2
\label{eq:dk11}
\end{eqnarray}
where $\phi_0$ is the value of the field at present epoch ($\tau=0$ or $t/t_0=1$).
The coefficients $\varepsilon_1$ and $\varepsilon_2$ depend on chosen value of $K$. 
From the analysis we find both these dependences  to be linear in $K$ and are given
by
\begin{eqnarray}
\varepsilon_1 (K) &=& -0.64 + 0.67 K \nonumber\\
\varepsilon_2 (K) &=& -0.80 + 0.45 K 
\label{eq:dk12}
\end{eqnarray}
The field $\phi$ thus has a linear dependence of diffusion parameter $K$.

\section{Conclusion}
\label{sec:conclusion}
In this work we have considered a model of diffusive dark matter and dark energy
where dark energy is represented by a  homogeneous $k-$essence scalar field  
$\phi$ with a  (non-canonical) Lagrangian with constant potential. We assume
particles of the dark matter fluid undergoing velocity diffusion
in the background medium of the $k-$essence scalar field $\phi$.
This diffusion establishes a dynamical relation between dark matter and dark energy
causing energy transfer from dark matter to the background
$k-$essence field. Motivations behind considering such models are to account for  
coincidence of measured values present day dark energy and dark matter densities.
We have shown that this model is supported by luminosity distance
and redshift data from SNe Ia observations.
The energy transfer between dark matter and background $k$-essence
field $\phi$ representing dark energy  shows up as a source term
in the continuity equations for the individual fluids of dark matter and dark energy.
The source term is proportional to a diffusion coefficient
which is a measure of average energy transfer per unit time
from dark matter to dark energy. In this work we realise this quantity
in terms of a dimension diffusion parameter $k$ introduced in Eq.\ \eqref{eq:c5}.
Using  time dependences of energy density and pressure
of the dark fluid as extracted 
from SNe Ia observations and the result that dark matter contributes
$\sim 27\%$ of the total energy density of present universe,
we obtain a constrain on the value of the diffusion parameter $K$
as $0 \leqslant K <0.44$.\\

Temporal behaviour of dark matter and dark
energy densities in this diffusive interaction scenario 
for different  values of diffusion parameter ($K$) are shown in Fig.\ \ref{fig:3}.
$K=0$ corresponds to non-interacting dark matter and dark energy. Then
Energy density of dark matter that satisfies the continuity
equation $\dot{\rho}_{\rm dm}  + 3H\rho_{\rm dm} = 0$ and 
$\rho_{\rm dm} \sim a^{-3}$. Non-zero (positive)
values of $K$ corresponds to energy transfer from dark matter
to the dark energy field $\phi$ and temporal behaviour of $\rho_{\rm dm}$ 
differs from that of $a^{-3}$. Higher values of $K$ 
corresponds to higher average value of transferred energy
from dark matter fluid to dark energy fluid.
At any given epoch, dark matter energy density is therefore
always lower for higher values of $K$. Plots of Fig.\ \ref{fig:3} 
depict this feature.\\

We assume the $k$-essence scalar field $\phi$ representing dark energy  to be homogeneous
and described by a non-canonical Lagrangian with constant potential which ensures
existence of a scaling relation Eq.\ (\ref{eq:dk3}). Using the scaling relation 
we described a methodology to obtain the temporal behaviour of the field $\phi$
without prior knowledge of  function $F(X)$ in $K-$essence Lagrangian.
The dependence of the field $\phi$ on the diffusion parameter $K$ is also obtained.
The obtained temporal behaviour of the field is expressed as
$\phi(t/t_0) = \phi_0 + \varepsilon_1 (t/t_0 - 1) + \varepsilon_2 (t/t_0 -1)^2$.
where $t_0$ is the time corresponding to present epoch. The coefficients
$\varepsilon_1$ and $\varepsilon_2$ are functions of diffusion parameter $K$ and 
the obtained dependence is expressed in Eq.\ \eqref{eq:dk12}.
We observe that SNe Ia data do not allow any room for 
terms of $t$ having order more than 2, independent 
of the values of diffusion parameter $K$. Thus the k−essence
field $\phi$ has similar temporal behaviour as that of quintessence
scalar field which is responsible for homogeneous inflation.

\end{document}